\documentstyle[11pt,epsfig]{article}

\def\be{\begin{eqnarray}}
\def\ee{\end{eqnarray}}
\def\ben{\begin{equation}}
\def\een{\end{equation}}

\def\calO{{\cal O}}
\def\calM{{\cal M}}

\newcommand{\e}{{\mbox{e}}}
\def\mN{m_N}
\def\del{\partial}
\def\vj{{\vec j}}
\def\vr{{\vec r}}

\def\vq{{\vec q}}
\def\vk{{\vec k}}
\def\vp{{\vec p}}

\def\vbp{{\vec {\bar p}}}
\def\vt{{\vec \tau}}
\def\vs{{\vec \sigma}}

\def\hatc{{\hat c}}
\def\hatr{{\hat r}}

\def\roughly#1{\mathrel{\raise.3ex\hbox{$#1$\kern-.75em%
\lower1ex\hbox{$\sim$}}}}
\def\lsim{\roughly<}

\def\MeV{{\mbox{MeV}}}

\def\nlo#1{\mbox{N$^{#1}$LO}}
\def\vA{{\vec A}}
\def\vV{{\vec V}}
\def\vtau{{\vec \tau}}

\def\vx{{\vec x}}

\def\Max{{\rm max}}
\def\MeV{{\mbox{MeV}}}
\def\GeV{{\mbox{GeV}}}

\def\mpit{{\tilde m}_\pi}

\renewcommand{\thefootnote}{\fnsymbol{footnote}}
\def\dR{{\hat d}^R}

\begin{document}

\hfill{USC(NT)-Report-01-9}

%\noindent \hfill \version

\vskip 10mm
\begin{center}

{\LARGE \bf
The \mbox{$\mu^- d$} capture rate in effective field theory }

\vskip 7mm

{\Large  
S. Ando\footnote{E-mail address : sando@nuc003.psc.sc.edu}, 
T.-S. Park\footnote{E-mail address : tspark@nuc003.psc.sc.edu}, 
K. Kubodera\footnote{E-mail address : kubodera@sc.edu}, 
F. Myhrer\footnote{E-mail address : myhrer@sc.edu}}

\vskip 3mm

{\Large \it Department of Physics and Astronomy, \\
University of South Carolina, Columbia, SC 29208, U.S.A. }

\end{center}

\vskip 7mm

%\centerline{\bf Abstract}\vskip 0.1cm 

Muon capture on the deuteron is studied
in heavy baryon chiral perturbation theory (HB$\chi$PT).
%up to ${\cal O}(Q^3)$ order. 
It is found that by far the dominant contribution
to $\mu d$ capture comes from a region 
of the final three-body phase-space 
in which the energy of the two neutrons
is sufficiently small for HB$\chi$PT 
to be applicable. 
The single unknown low-energy constant
having been fixed from the tritium beta decay rate, 
our calculation contains no free parameter.
Our estimate of the $\mu d$ capture rate
is consistent with the existing data.
The relation between $\mu d$ capture  
and the $\nu d$ reactions, which are
important for the SNO experiments,
is briefly discussed.

\vskip 1cm

\noindent
PACS : 12.39.Fe, 23.40.-s

\renewcommand{\thefootnote}{\#\arabic{footnote}}
\setcounter{footnote}{0}

\newpage
\section{Introduction} 

Electroweak processes in the two-nucleon systems
invite detailed studies for multiple reasons.
From the nuclear physics point of view,
these processes offer a valuable testing ground 
of the basic inputs of nuclear physics.
In astrophysics, the precise knowledge of
the $pp$ fusion cross section
is of crucial importance for building
a reliable model for stellar evolution
\cite{bm69}.
Furthermore, experiments
at the Sudbury Neutrino Observatory (SNO)
\cite{SNO,ahmetal01} 
to observe solar neutrinos 
with a heavy-water Cerenkov counter
have made it extremely important 
to estimate the $\nu d$ reaction cross sections
with high precision.

In this note we study 
$\mu d$ capture: 
$\mu^- + d \rightarrow \nu_\mu + n +n$,
using effective field theory (EFT).
Our work is connected to the above-mentioned
urgent need of accurate estimates of the 
$\nu d$ cross sections, $\sigma_{\nu d}$.
To expound this connection,
we will first explain the 
standard nuclear physics approach (SNPA), 
see {\it e.g.} Ref.\cite{cs98}. 
This is a highly successful method for describing
nuclear responses to electroweak probes. 
In this approach 
we consider one-body (1B) impulse approximation
terms and two-body (2B) exchange-current terms
acting on non-relativistic nuclear wave functions,
with the exchange currents derived from 
a one-boson exchange model. 
The vertices in the relevant Feynman diagrams
are obtained from a Lagrangian
constructed to satisfy the low-energy theorems 
and current algebra \cite{cr71},
while the nuclear wave functions 
are generated by solving the A-body Schr\"{o}dinger
equation, $H|\Psi_A\rangle\,=\,E|\Psi_A\rangle$,
where the Hamiltonian $H$ contains 
realistic phenomenological nuclear potentials.
The most elaborate study of $\mu d$ capture based on SNPA
was carried out by Tatara {\it et al.} (TKK) \cite{tkk}
and by Adam {\it et al.} \cite{adaetal90}. 

Now, the best available estimation of $\sigma_{\nu d}$
based on SNPA is due to Nakamura {\it et al.} (NSGK)
\cite{nsgk},
while that based on EFT is due to Butler, Chen, and Kong
(BCK) \cite{bck01}.
Since EFT is a general framework \cite{review}, 
it can give model-independent results,
{\underline {\it provided}}
all the low-energy coefficients (LEC)
in the effective Lagrangian, ${\cal{L}}_{eff}$,
are predetermined.
${\cal{L}}_{eff}$ considered by BCK, however, 
does contain one unknown parameter ($L_{1A}$), 
which they adjusted to
reproduce $\sigma_{\nu d}$ obtained by NSGK.
After this adjustment,
the results of BCK are found to be in perfect agreement
with those of NSGK.
The fact that an {\it ab initio} calculation
(modulo one free parameter) based on EFT
reproduces $\sigma_{\nu d}$ of NSGK extremely well
offers strong support to the calculation based on SNPA.
At the same time, it stresses the importance 
of carrying out an EFT calculation free from
an adjustable parameter.
An interesting possibility is to use $\mu d$ capture data
as input to control the unknown LEC.
An immediate question, however, is 
whether this process is ``gentle" enough
to be amenable to EFT.
The substantial energy transfer
accompanying the disappearance of a muon
can lead to a region of 
the final three-particle phase space
in which the intrinsic state of the two neutrons
receives such a large momentum
that the applicability of EFT becomes a delicate issue.
Let this unfavorable kinematical region be called
the ``dangerous" region.
The problem of the dangerous region
is reminiscent of the difficulty 
one encounters in applying EFT to 
threshold pion production in 
$N+N\rightarrow N+N+\pi$ \cite{bypark,Kolck3,sato,APM}.
It will turn out (see below),
however, that, unlike the pion production case,
$\mu d$ capture receives
only a tiny fraction of contribution
from the dangerous region,
and therefore the theoretical uncertainty caused
by the dangerous region is practically negligible.                      

The calculation in \cite{bck01} used the 
power divergence scheme (PDS) \cite{PDS}.
We employ here another form of EFT,
the Cut-off Weinberg scheme (CWS).
In CWS, the transition operators
are derived from irreducible diagrams
in heavy-baryon chiral perturbation theory
(HB$\chi$PT), and the nuclear matrix elements
are obtained by sandwiching them between
initial and final nuclear wave functions obtained in SNPA.
As discussed in \cite{pp-hep}, 
%a next-to-next-to-leading order (N$^2$LO) calculation
a next-to-next-to-next-to-leading order (\nlo3) calculation
based on CWS contains one unknown LEC,
denoted by $\hat{d}^R$.
Like $L_{1A}$ discussed by BCK, the parameter $\hat{d}^R$
controls the strength of a short-range exchange-current term
and, once $\hat{d}^R$ is fixed from data,
we can make a definite EFT
prediction for $\sigma_{\nu d}$.
We therefore investigate here 
the relation between $\hat{d}^R$
and the $\mu d$ capture rate,
$\Gamma_{\mu d}$.
Our study is essentially of exploratory nature,
given the present limited accuracy
(see below)
of the experimental value of $\Gamma_{\mu d}$. 

Another important point
concerning $\hat{d}^R$ is that,
as emphasized in \cite{pp-hep},
the strength of $\hat{d}^R$ can be reliably related to
the tritium $\beta$-decay rate, $\Gamma_\beta^t$.
Thus, using the experimental value 
of $\Gamma_\beta^t$, 
which is known with high precision,
one can determine $\hat{d}^R$
and then proceed to make predictions on
various two-nucleon weak-interaction processes,
including the $\mu d$ capture rate, 
pp fusion rate, and $\nu d $ cross sections.
We will present here the first estimate 
of the $\mu d$ capture rate
obtained in this approach;
the $pp$ fusion rate has already been discussed 
in \cite{pp-hep}, and the $\nu d$ cross sections
will be reported elsewhere \cite{nud01}.

\section{The capture rate}

Although $\mu d$ capture can in principle occur 
from the two $\mu d$ hyperfine states 
($S_{\mu d}=1/2$ and $S_{\mu d}=3/2$), 
the capture is known to 
take place practically uniquely
from the hyperfine doublet state.
Therefore, concentrating on this dominant capture,
we refer to hyperfine-doublet $\mu d$ capture
simply as $\mu d$ capture
and denote the hyperfine-doublet $\mu d$ capture rate
by $\Gamma_{\mu d}$.
The measured value of $\Gamma_{\mu d}$ is 
$\Gamma_{\mu d}^{exp}=409
 \pm 40$ s$^{-1}$ \cite{caretal89}
and
$\Gamma_{\mu d}^{exp}=470 
\pm 29$ s$^{-1}$ \cite{baretal86}. 
We remark that 
a high-precision measurement of $\Gamma_{\mu d}$
is being contemplated at PSI \cite{kametal}.

Denoting by $L$ the orbital angular momentum
of the two-neutron relative motion in the final state,
we can write
\ben
\Gamma_{\mu d}=\sum_{L=0,1,2 ...}\Gamma_{\mu d}^L,
\een
where $\Gamma_{\mu d}^L$ is 
the rate of $\mu d$ capture leading to the $L$ state.
Here we shall be primarily concerned with
$\Gamma_{\mu d}^{L=0}$
since it is this quantity
that contains information about $\hat{d}^R$.
The contributions of $\Gamma_{\mu d}^L$ ($L\ge 1$)
are significant,\footnote{For example, 
$\Gamma_{\mu d}^{L=1}\approx\frac{1}{3}\Gamma_{\mu d}$
according to Ref.\cite{tkk}.}
but their calculation is not expected to involve 
any major EFT-related issues.
In general, due to the centrifugal force, 
the $L \ge 1$ contributions cannot be 
too sensitive to short-range physics,
which implies that chiral expansion 
for them should converge rapidly.
Specifically, the $L=1$ contributions are dominated by the
axial-charge (AC) and E1 transitions,
whose one-body operators 
(which are NLO in chiral counting)
are well-known. 
The lowest order 
meson-exchange corrections (MEC) 
to the one-body operators 
come from soft one-pion-exchange (OPE), which is 
\nlo2 in chiral counting.
These soft-OPE terms, dictated by chiral symmetry,
are well known, and 
they are model-independent.
For $L\ge 2$ states, 
within the accuracy of our evaluation 
only 
one-body contributions \cite{tkk} have to be included.
In our exploratory study, therefore,
we concentrate on a detailed evaluation of
$\Gamma_{\mu d}^{L=0}$, and  
for $\Gamma_{\mu d}^L$ ($L\ge 1$)
we simply use the results obtained by
Ref.\cite{tkk}. 

Muon capture by the deuteron is effectively described by 
the current-current hamiltonian of weak interactions
\be
H_W = \frac{G_V}{\sqrt{2}} \int d^3x\,
 L_\alpha(\vx) J^{\alpha}(\vx)
 + \mbox{h.c.} ,
\ee
where the leptonic and the hadronic charged currents are 
\be
L_\alpha(\vx)&=&
\bar\psi_\nu(\vx) \gamma_\alpha (1-\gamma_5) \psi_\mu(\vx)  
\; \; \; \; \;{\rm and} 
\nonumber \\
J^{\alpha}(\vx) &=& (V^\alpha - A^\alpha)^{a=1}(\vx) 
 - i (V^\alpha - A^\alpha)^{a=2}(\vx),
\ee
respectively, and $G_V= 1.14939 \times 10^{-5}\ \GeV^{-2}$ 
\cite{hardy};
$\alpha$ ($a$) is the Lorentz (isospin) index.
In the center-of-mass system of the initial $\mu$-$d$ atom
from which capture occurs, we can safely assume
$\vp_\mu=\vp_d={\vec 0}$.
Consequently, the four-momentum transfer 
to the leptonic system,
$q^\alpha \equiv (p_\nu - p_\mu)^\alpha$, reads
$(q^0,\ \vq) =(E_\nu - m_\mu,\ \vec{p}_\nu)$. 
The $\mu$-$d$ capture amplitude is then given by 
\be
\langle f |H_W| i \rangle
= \frac{G_V}{\sqrt{2}} \; \Psi_{\mu-d}({\vec 0}) \; 
l_\alpha \; 
 \langle 
  \Psi_{nn}(-\vq , \vp;s_1 s_2)
  | j^{\alpha}(\vq) |
  \Psi_d(s_d) 
  \rangle ,
\label{eq;amplitude}
\ee
where the Fourier transformed currents are
\be
j^{\alpha}(\vq )&\equiv& 
 \int d^3\vx\,\e^{-i\vq \cdot \vx}\, J^{\alpha}(\vx),
\\
l_\alpha &\equiv&
 \e^{i \vx \cdot \vq } 
\langle \nu (p_\nu,s_\nu) | L_\alpha(\vx) |\mu^- (p_\mu,s_\mu)\rangle
= {\bar u}_\nu(\vp_\nu,s_\nu) \gamma_\alpha (1-\gamma_5) 
 u_\mu(\vp_\mu,s_\mu).
\ee 
In Eq.(\ref{eq;amplitude}),
$\Psi_{\mu-d}(\vec 0)$ = 
$ 1 / \sqrt{\pi} a_0^{\frac32}$
is the $1S$ wave function of the $\mu$-$d$ atom, 
where $a_0 \equiv  (m_\mu + m_d) / m_\mu m_d \alpha$, 
with 
$\alpha \simeq 1/137.036$ 
the fine structure constant.\footnote{
The small correction due to the finite size of
the deuteron is taken into account in the actual calculation.}
$\Psi_d(s_d)$ in Eq.(\ref{eq;amplitude})
represents the deuteron wave function 
with the z-component of its spin $s_d$; 
$\Psi_{nn}(-\vq ,\vp;s_1 s_2)$
represents the final $nn$ wave function,   
with total $nn$ momentum $-\vq $,  
relative $nn$ momentum $\vec{p}$ = $(\vp_1 - \vp_2)/2$,
and the z-components of the neutron spins, $s_1$ and $s_2$.  
It is easy to obtain 
\be
\Gamma_{\mu d}
 &=& \frac{|G_V \Psi_\mu({\vec 0})|^2}{4(2 J_{\mu d} + 1)}
 \int \frac{d^3\vp}{(2\pi)^3}
 \int \frac{d^3\vp_\nu}{(2\pi)^3}
 \, 2\pi \delta(\Delta E)
\nonumber \\
&& 
  \sum_{S_{\mu_d}=-J_{\mu d}}^{J_{\mu d}}
  \sum_{s_1 s_2 s_\mu s_d} \Bigg|
  \langle \Psi_{nn} | j^\alpha(\vq) | \Psi_d\rangle
   l_\alpha\,
   \langle \frac12,s_\mu;\,1, s_d | J_{\mu d}, S_{\mu d}\rangle
   \Bigg|^2 ,
\label{cap}
\ee
where $J_{\mu d}=\frac12$,
$\Delta E$ is the energy difference between the
final and initial states,  
$\Psi_d\equiv \Psi_d(s_d)$ and 
$\Psi_{nn}\equiv \Psi_{nn}(-\vq ,\vp;s_1 s_2)$.

\section{HB$\chi$PT Lagrangian and the hadronic currents}
\indent
To calculate the capture rate 
we adopt Weinberg's power counting rule.
In HB$\chi$PT the leading order (LO) Lagrangian is given by
\be
{\cal L}_0 &=& {\bar B}\left[ i v\cdot D
+ 2 i g_A S\cdot \Delta \right] B
- \frac{1}{2} \sum_A C_A \left({\bar B} \Gamma_A B\right)^2
+ f_\pi^2 {\rm Tr}\left(i \Delta^\mu i \Delta_\mu\right)
+ \frac{f_\pi^2}{4} {\rm Tr}(\chi_+)
\label{chiralag2} \ee
with
$D_\mu \equiv \del_\mu + 
 \frac{1}{2} \left[\xi^\dagger,\, \del_\mu \xi\right]
-\frac{i}{2}\xi^\dagger {\cal R}_\mu \xi - \frac{i}{2}
\xi {\cal L}_\mu\xi^\dagger$,
$\Delta_\mu = \frac{1}{2} \left\{\xi^\dagger,\, \del_\mu \xi\right\}
+\frac{i}{2}\xi^\dagger {\cal R}_\mu \xi - \frac{i}{2}
\xi {\cal L}_\mu\xi^\dagger$
and
$\chi_+ = \xi^\dagger \chi \xi^\dagger + \xi \chi^\dagger \xi$,
where
${\cal R}_\mu = \frac{\tau^a}{2} \left(
  {\cal V}^a_\mu + {\cal A}^a_\nu\right)$
and
${\cal L}_\mu = \frac{\tau^a}{2} \left(
  {\cal V}^a_\mu - {\cal A}^a_\nu\right)$
denote the external gauge fields.
In the absence of the external scalar
and pseudo-scalar fields $\chi=m_\pi^2$, 
and we define the pion field as 
$\xi = {\rm exp}\left(i\frac{{\vec \tau}\cdot {\vec \pi}}{2 f_\pi}\right)$.
It is convenient to choose  
the four-velocity $v^{\mu}$ and the spin operator $S^{\mu}$ 
as 
$v^\mu = (1,\, {\vec 0})$
and
$S^\mu = \left(0,\, \frac{{\vec \sigma}}{2}\right)$. 

The next-to-leading-order (NLO) Lagrangian 
(including the ``$1/m_N$" terms) in
the one-nucleon sector is given in \cite{csNLO}
while that in the two-nucleon sector is given in 
\cite{Kolck3}\footnote{
Our definition of the pion field differs from that used in Ref.
\cite{Kolck3} by a minus sign. 
}. 
Combining them, we can write
the NLO Lagrangian relevant to our case as  
\be
{\cal L}_1 &=& {\bar B} \left(
  \frac{v^\mu v^\nu - g^{\mu\nu}}{2 \mN} D_\mu D_\nu
   + c_1 \mbox{Tr} \chi_+
   + \left(4 c_2 - \frac{g_A^2}{2 \mN}\right) (v\cdot i\Delta)^2 
   + 4 c_3 \; i\Delta\cdot i\Delta + 
\right. 
\nonumber \\ 
&&\left. \ \ \  + \left(2 c_4 + \frac{1}{2\mN}\right)
  \left[S^\mu, \, S^\nu \right] 
\left[ i \Delta_\mu , \, i\Delta_\nu \right]
- \; i \; \frac{1+c_6}{\mN}
  \left[ S^\mu, \,S^\nu \right] f_{\mu\nu}^+ \right) B 
\nonumber \\
 && -\ 4 i d_1 \,
 {\bar B} S\cdot \Delta B\, {\bar B} B
 + 2 i d_2 \,
  \epsilon^{abc}\,\epsilon_{\mu\nu\lambda\delta} \; 
v^\mu \Delta^{\nu,a}
 {\bar B} S^\lambda \tau^b B\, {\bar B} S^\delta \tau^c B
+ \cdots, 
\label{Lag1}\ee
where
$\epsilon_{0123}=1$, $\Delta_\mu = \frac{\tau^a}{2} \Delta^a_\mu$,
and 
$f_{\mu\nu}^+$ =
 $\xi (\del_\mu {\cal L}_\nu - \del_\nu {\cal L}_\mu
   - i \left[{\cal L}_\mu,\,{\cal L}_\nu\right]) \xi^\dagger $
 + $\xi^\dagger (\del_\mu {\cal R}_\nu - \del_\nu {\cal R}_\mu
   - i \left[{\cal R}_\mu,\,{\cal R}_\nu\right]) \xi$. 
We find it convenient to use
the dimensionless low-energy constants
$\hat c$'s and $\hat d$'s defined by
\be
c_{1,2,3,4} = \frac{1}{\mN}\, \hat c_{1,2,3,4},
\ \ \ {\rm and} \ \ \ 
d_{1,2} = \frac{g_A}{m_N f_\pi^2}\,\hat d_{1,2}.
\ee
The values of these low energy constants, $\hatc_{1,2,3,4}$, 
are taken from Ref.\cite{csNLO}:
\be
\hat c_1 = -0.60 \pm 0.13,
\ \ \
\hat c_2 = 1.67 \pm 0.09,
\ \ \
\hat c_3 = -3.66 \pm 0.08,
\ \ \
\hat c_4 = 2.11 \pm 0.08\,
\ee
and $c_6=\kappa_V=3.70$.
These values were
determined at tree level (or NLO) in the one-nucleon sector,
which correspond to \nlo3
in our two-nucleon calculation.

\subsection{The one-body currents}
\indent

The one-body currents can be 
obtained either by an explicit HB$\chi$PT calculation
or by the Foldy Wouthuysen (FW) reduction of the
well-known relativistic expressions.
The former method requires one-loop diagrams
which consist of vertices from ${\cal L}_0$
(for \nlo2 contributions)
and those which contain one vertex from ${\cal L}_1$
(for \nlo3 contributions);
also needed are the corresponding counter-terms from
${\cal L}_2$ and ${\cal L}_3$.
We adopt here the FW reduction method for convenience.
Since the range of 
$t \equiv q^2 = m_\mu (m_\mu-2 E_\nu)$
for $\mu d$ capture is small 
($-m_\mu^2 \lsim t \le m_\mu^2$),
the $t$-dependences
in the standard form factors, 
$F_{1,2}^V(t)$ and $G_{A}(t)$,
give only less than 2 \% effects,
which can be reliably taken into account 
by expansion in $t$,
$F_1^V(t)$ = $1 + \frac{t}{6} r_V^2 + {\cal O}(t^2)$,
$F_2^V(t)=\kappa_V + {\cal O}(t)$
and 
$G_A(t)$ = $g_A\left(1 + \frac{t}{6} r_A^2 + 
{\cal O}(t^2)\right)$. 
Keeping the terms linear in $t$ 
is consistent with HB$\chi$PT to the order we calculate. 
Some caution is required for the $G_P(t)$ term,
which contains the pion-pole contributions,
\be
\frac{G_P(t)}{2 m_N} \equiv
 \beta(t) \, \frac{2 m_N G_A(t)}{m_\pi^2 - t}, 
\label{eq:GP}
\ee
where $\beta(t)$ is a slowly varying function. 
Comparison with the explicit HB$\chi$PT calculation 
up to \nlo2 \cite{gp} leads to
\be
\beta(t)= \frac{f_\pi g_{\pi NN}}{g_A m_N}
 - \frac16 r_A^2 m_\pi^2
 + {\cal O}\left(\frac{Q^3}{\Lambda_\chi^3}\right)
= 1 + \left[(-2.0 \sim 1.5) \pm 0.3\right] \ \%\,.
\ee
The authors of Ref. \cite{tkk} found that
$\Lambda_{\mu d}$ is reduced
only by $\sim$2 \% when $\beta$ increases by 10\ \%.
Thus, 
limiting ourselves to the $\beta=1$ case 
entails at most $0.4\ \%$ error.\footnote{ 
\protect
If the deviation of $\beta$ from 1
were important,
we would have to include the 
one-body pseudoscalar term,
$\hat P_{\rm 1B}$ in Eq.(\ref{1BPCAC}),
as is necessary for the two-body 
$\hat P_{\rm 2B}$ = $\hat P$ in Eq.(\ref{PCAC}).}

The resulting one-body vector (1B) current components are
\be
V^{0,-}_{\rm 1B}(\vq) &=& \sum_i
\tau_i^- \;  \e^{-i \vq\cdot \vr_i} \left[
 1 + \frac{t}{6} r_V^2
- \frac{\vq^2}{8m_N^2} 
 + (1+2\kappa_V) \frac{i \vq \cdot \vs_i \times \vbp_i}{4 m_N^2}
 - \kappa_V \frac{\vq^2}{4 m_N^2}\right],
\nonumber \\
\vV^-_{\rm 1B}(\vq) &=& \sum_i 
\tau_i^- \e^{-i \vq\cdot \vr_i} \left[
 \frac{\vbp_i+\frac{i}{2}(1+\kappa_V)\vq\times\vs_i}{m_N}
 + (1+2\kappa_V)
  \left(i \vs_i\times \vbp_i-\frac12 \vq\right) \frac{\omega}{4 m_N^2}
 \right]
\ee
where $\tau_i^- \equiv \frac12 (\tau_i^x - i \tau_i^y)$
and ${\vec {\bar p}}_i = (\vp_i^{\; \prime} + \vp_i )/2$.
The one-body axial-vector current components can be 
written for convenience as
\be
A^\alpha_{\rm 1B} = \hat A^\alpha_{\rm 1B} + 
\frac{q^\alpha}{m_\pi^2 - t} 
  q_\beta \hat A^\beta_{{\rm 1B}},
\label{1BPCAC} 
\ee
which defines $\hat A_{\rm 1B}$, where 
\be
\hat A^{0,-}_{\rm 1B}(\vq) &=& \sum_i
 g_A \; \tau_i^- \; \e^{-i \vq\cdot \vr_i} \left[
 \frac{\vs_i\cdot \vbp_i}{m_N}
 - \omega \frac{\vs_i\cdot\vq}{8 m_N^2}
\right],
\nonumber\\
\hat \vA^-_{\rm 1B}(\vq) &=& \sum_i 
 g_A \tau_i^- \e^{-i \vq\cdot \vr_i} \left[
 \vs_i\left(1+\frac{t}{6} r_A^2\right) 
 + \frac{2 (\vbp_i\, \vs_i \cdot \vbp_i - \vs_i \, \vbp_i^2)
     - \frac12 \vq\,\vs_i\cdot\vq
     + i \vq\times \vbp_i}{4 m_N^2}
\right],
\ee
The above equations correspond to 
\nlo2 in HB$\chi$PT \cite{amk}.
Apart from the mentioned $t$-dependence 
of the form factors,
the \nlo2 contribution is found to be negligible,
$\sim 0.1$~\%, indicating a rapid convergence.
We therefore limit ourselves to \nlo2 
for the one-body currents
although, to be completely consistent,
we should in principle include \nlo3 . 
%which requires considerable amount of work. 
%
%In principle, we must go up to \nlo3 to be completely
%consistent, which requires considerable amount of work. 
%However, as we shall see,
%(about 1~\%),
%and therefore up to \nlo2 it should be sufficient
%to include the one-body currents alone.
%However, it has been found that the \nlo2 contribution
%other than the above $t$-dependence is negligible,
%$\sim 0.1\ \%$.

\subsection{Two-body exchange currents}
\indent

Since the evaluation of the two-body exchange current is
the focus of this work we discuss  the 
various parts of this exchange current in detail. 
We write the two-body vector and axial-vector currents
%in momentum space 
as  
$V_{\rm 2B}^\mu(\vk_1 ,\vk_2)$
and $A_{\rm 2B}^\mu(\vk_1 ,\vk_2)$,
where $\vk_i = \vp_i^{\; \prime} - \vp_i$ is the momentum
transferred to the $i$-th nucleon.
For the two-body vector-charge current 
we know  
$V^0_{\rm 2B}(\vk_1,\vk_2) = {\cal O}(Q^4)$ \cite{V0},
which therefore can be ignored.  
%to the order of our calculation.
%We therefore can ignore $V^0_{\rm 2B}(\vk_1,\vk_2)$. 
The spatial components of the vector current have a 
one-pion-exchange contribution of order $Q^2$ \cite{M1}: 
\be
{\vec V}_{\rm 2B}(\vk_1 ,\vk_2) &=& -i (\tau_1\times\tau_2)^- \; 
\frac{g_A^2}{4 f_\pi^2} 
\bigg[
\frac{\vs_1 \, (\vs_2\cdot\vk_2)}{m_\pi^2 - k_2^2}
- \frac{\vs_2 \, (\vs_1\cdot\vk_1)}{m_\pi^2- k_1^2} 
\nonumber \\ 
&& 
+ \left(\frac{ \vs_1\cdot\vk_1}{m_\pi^2 - k_1^2}\right)
 \left(\frac{\vs_2\cdot\vk_2}{m_\pi^2 - k_2^2} \right)
 (\vk_2 - \vk_1)  
\bigg]
  + {\cal O}(Q^4),
\ee
where 
$f_\pi \simeq 93$ MeV is the pion decay constant,  
and we make use of the notation
$(\tau_1 \times \tau_2)^- \equiv 
(\tau_1 \times \tau_2)^x - i (\tau_1 \times \tau_2)^y$. 
%where $\otimes=\pm,\times$.
%and we make use of the notation: 
%\be
%(\tau_1 \otimes \tau_2)^- \equiv 
%(\tau_1 \otimes \tau_2)^x - i (\tau_1 \otimes \tau_2)^y,
%\ \ \
%\ \ {\rm where} \ \
%\otimes=\pm,\times.
%\ee 
Analogously to Eq.(\ref{1BPCAC}), 
we write the axial-vector current as \cite{ACGT,pp-hep},
\be
A^\alpha_{\rm 2B} = \hat A^\alpha_{\rm 2B} + 
\frac{q^\alpha}{m_\pi^2 - t} 
  \left( q_\beta \hat A^\beta_{{\rm 2B}} + 
\hat P \right),
\label{PCAC} 
\ee
with
\be
\hat A^{0}_{\rm 2B}(\vk_1 ,\vk_2) &=&
  i\, (\vt_1 \times \vt_2)^- \, 
\left(\frac{g_A}{4 f_\pi^2}\right) 
\left(\frac{{\vec \sigma}_1\cdot\vk_1}{m_\pi^2 - k_1^2 }\right)
- \frac{2 g_A}{m_N f_\pi^2} 
\left(\hat c_2 + \hat c_3 - \frac{g_A^2}{8}\right)
\tau_1^- \frac{k_1^0\,\vs_1\cdot\vk_1}{m_\pi^2-k_1^2}
\nonumber \\
&&+ (1 \leftrightarrow 2),
\\
\hat \vA_{\rm 2B}(\vk_1 ,\vk_2)
 &=& - \frac{g_A}{2 m_N f_\pi^2} \left\{\left[
 \frac{i}{2} (\vtau_1\times \vtau_2)^- \; {\vec {\bar p}}_1
 + 4 \hat c_3 \tau_2^- \; \vk_2
 + (\hat c_4 + \frac14) (\vtau_1\times \vtau_2)^- 
(\vs_1\times \vk_2) 
 \right.
 \right.
\nonumber \\
&&\left.\ \ \
 +\ \left(\frac{1+c_6}{4}\right)
  (\vtau_1\times \vtau_2)^- (\vs_1\times \vq )
 \right] \frac{\vs_2\cdot\vk_2}{m_\pi^2 - k_2^2}
\nonumber \\ &&\left. 
 +
 \left[ 2 \hat d_1 (\tau_1^- \; \vs_1 + \tau_2^- \; \vs_2)
 + \hat d_2  (\vtau_1\times \vtau_2)^- (\vs_1\times \vs_2)
 \right]
 + (1\leftrightarrow 2)
\frac{}{} \right\}\, ,
\\
\hat P (\vk_1 , \vk_2)
 &=& - \frac{g_A m_\pi^2}{2m_N f_\pi^2} \left\{
  8\hat c_1 \vtau_2^{\; -} 
 \left(\frac{\vs_2\cdot\vk_2}{m_\pi^2- k_2^2}\right)
 + (1\leftrightarrow 2)\right\} \, .
\ee 
Only one combination of the LEC, 
$\hat d_1$ and $\hat d_2$, is relevant for 
the $\mu d$ capture process,
\be
\hat d^R \equiv \hat d_1 +2 \hat d_2 +
\frac13 \hat c_3 + \frac23 \hat c_4 + \frac16.
\label{dR}\ee
Exactly the same combination of LEC's 
appears in triton $\beta$-decay,
$pp$-fusion and the solar $hep$ process\cite{pp-hep}. 
Adopting the same strategy as 
in Ref. \cite{pp-hep}, 
we fix $\dR$ from $\Gamma_\beta^t({\rm exp})$,
the experimental value of
the tritium $\beta$-decay rate.

To facilitate the calculations, we perform a
Fourier transformation (FT) 
of the above two-body currents. 
To control short-range physics 
in performing FT, we introduce
a Gaussian cut-off regulator
\be
S_\Lambda(\vk^2)
=   \exp\left(-\frac{\vk^2}{2\Lambda^2}\right) .  
\label{regulator} 
\ee
where $\Lambda$ is a cut-off parameter.
It is to be emphasized
that, although our calculation without regularization
involves no infinities,
we still need a regulator 
since EFT, by definition,
is valid only up to a certain momentum scale.
The regulated delta and Yukawa functions read
\be
\delta_\Lambda^{(3)}(\vr) &\equiv&
 \int \!\frac{d^3\vq}{(2\pi)^3}\,
  S_\Lambda^2(\vq^2)\, \e^{ i \vq\cdot \vr} 
   = \frac{\Lambda^3}{(4\pi)^{\frac32}}
 \, \exp\left(-\frac{\Lambda^2 r^2}{4}\right),
\label{regdelta}
\nonumber \\
y_{0\Lambda}(m,r) &\equiv&
 \int \!\frac{d^3\vq}{(2\pi)^3}\,
  S_\Lambda^2(\vq^2)\, \e^{ i \vq\cdot \vr}\,
  \frac{1}{\vq^2 + m^2}\,,
\ee
We remark that this is exactly 
the same regularization method as used in Ref. \cite{pp-hep}.

In performing FT, we need to specify
the time components of the momentum 
transferred to the nucleons.
Energy conservation imposes the constraint: 
$ k_1^0 + k_2^0 = - q^0 = m_\mu - E_\nu $.
In our calculation
we will adopt the so-called
fixed-kinematics assumption (FKA) \cite{bypark}, 
where the energy transfer is assumed 
to be shared equally between the two nucleons, i.e. 
$ k_1^0= k_2^0 $ = $ \frac{m_\mu - E_\nu}{2} $,
which naturally brings in 
the quantity %$\mpit$
$\mpit\equiv \sqrt{m_\pi^2 - (m_\mu-E_\nu)^2/4}$.
The uncertainty related to FKA becomes large
as $|q^0|$ grows.  
The contribution from the large $|q^0|$ region,
however, will turn out to be so tiny that
the assumptions related to $k_i^0$ cause
little uncertainty in our calculation.

\section{The capture rate for the transition 
to the ${}^1S_0$  $nn$ state}
\indent

The deuteron and the ${}^1S_0$ wave function may be written as
\be
\psi_d(\vr; s_d) = \frac{1}{\sqrt{4 \pi} r} \left[
 u_d(r) + \frac{S_{12}(\hat r)}{\sqrt{8}} w_d(r) \right]
 \chi_{1,s_d} \xi_{0,0},\ \ \ 
\psi_0(r) = \frac{1}{\sqrt{4 \pi} r} u_0(r) \chi_{0,0} \; \xi_{1,-1}
\ee
with
$\int_0^\infty dr\, \left[u_d^2(r) + w_d^2(r) \right] = 1$
and
$ \lim_{r\rightarrow \infty} u_0(r)
= \frac{\sin\delta_0}{p} 
  \left[ \cos pr + \cot\delta_0 \sin pr \right]$. 
Here 
$S_{12}(\hat r)= 3 \vs_1\cdot \hatr \,\vs_2 \cdot \hat r -
\vs_1\cdot\vs_2$, 
$\chi$ ($\xi$) is the Pauli spinor (isospinor),
and
$\delta_0$ is the $nn$ ${}^1S_0$ phase shift.
To facilitate numerical work, we approximate 
$\Delta E$ as 
\be
\Delta E 
= E_\nu 
 + 2 \sqrt{m^2 + \vp^2  + \frac{E_\nu^2}{4}}
 - M_{\mu d}
 + {\cal O}\left(\frac{(\vp_\nu\cdot\vp)^2}{4 m^3}\right) , 
\ee
where $m\equiv m_n = 939.566\ \MeV$ is the neutron mass,
$M_{\mu d}\equiv m_\mu+m_d = 1981.272\ \MeV$.
In our calculation we will neglect the
${\cal O}\left(\frac{(\vp_\nu\cdot\vp)^2}{4 m^3}\right)$ term 
since, as we shall show, 
the major contributions comes from the low
$p\equiv |\vp |$ region.
\def\dlz{\delta_{\lambda,0}}
\def\bGP{{\bar G}_P}
Choosing the $z$-axis along $\vp_\nu$, we write
$\vq= \vp_\nu = E_\nu\,\hat z$.
This simplifies the structure of 
the transition amplitudes as
\be
\langle \psi_0| j^0(\vq) | \Psi_d(s_d)\rangle = \delta_{s_d,0} \calM_t,
\ \ \ 
\langle \psi_0| \hat e_\lambda^* \cdot \vj(\vq) | \Psi_d(s_d)\rangle
= \delta_{s_d,\lambda} \calM_\lambda,
\ee
where 
$ \hat e_\pm = \mp (\hat x \pm i \hat y)/\sqrt{2}$, 
$\hat e_0 = \hat z$,
and $\lambda=\pm 1, 0$.
We decompose the matrix elements
into vector and axial vector current contributions,
${\calM}_{t,\lambda}= 
{\calM}_{t,\lambda}[V] - {\calM}_{t,\lambda}[A]$, and 
arrive at
\be
\Gamma_{\mu d}^{L=0}
 &=& \frac{|G_V \Psi_\mu({\vec 0})|^2}{2 \pi^2}
 \int_0^{p^\Max} d p\, 2 p^2 E_\nu^2 
  \left(1 - \frac{E_\nu}{M_{\mu d}}\right)\, 
 \frac23 \Bigl|2 \calM_{-1}+\calM_{0} - \calM_t\Bigr|^2\,.
\ee
Note that
$\calM_{-1}= - (\calM_{+1}[V]+\calM_{+1}[A])$,
$\calM_{0}= - \calM_{0}[A]$
and, to the order under consideration,
$\calM_{t}= - \calM_{t}[A]$.
The matrix elements of the 
vector current are
\be
\calM_\lambda[V] &=&
 \lambda\sqrt2\int_0^\infty\! dr\,\left\{
   q \; u_0\left(u_d \; j_0 - \frac{j_2 \; w_d }{\sqrt2} \right)
     \frac{\mu_V}{2 m_N}
   - \omega \; u_0^{(1)}
      \left(u_d + \frac{w_d}{\sqrt2}\right) j_1 
       \frac{2 \mu_V-1}{4 m_N^2}
  \right\}
\nonumber \\
&+& \lambda \,
(4\sqrt2) \left(-\frac{g_A^2}{8 f_\pi^2} \right) q
 \int_0^\infty\! dr \,u_0 \int_{-\frac12}^{\frac12}\!dx\,
 \left[
 \left(j_0^x \; u_d - \frac{j_2^x \; w_d}{\sqrt2}\right)
  \left(y_0^L - \frac23 y_1^L\right)
\right.
\nonumber \\
&&\left.
  - x q r j_1^x \left(u_d + \frac{w_d}{\sqrt2}\right) y_0^L
 + \frac13 \left( j_2^x \; u_d - 
    \left(\sqrt2 j_0^x + \frac{j_2^x}{\sqrt2}\right) w_d\right) y_1^L
 \right] \, ,
\ee
where
$j_n^x\equiv j_n(q r x)$ 
are the spherical Bessel functions, 
$ y_n^L\equiv y_{n\Lambda}\left(\sqrt{m_\pi^2 + 
\frac{1-4 x^2}{4} \vq^2},r\right)$ 
and 
$y_{1\Lambda}(m,r) 
\equiv - r \frac{\del}{\del r} y_{0\Lambda}(m,r)$
and
$y_{2\Lambda}(m,r) \equiv \frac{1}{m^2}
  r \frac{\del}{\del r} \frac{1}{r}
  \frac{\del}{\del r} y_{0\Lambda}(m,r)$.
 
Using Eq.(\ref{PCAC}),
we obtain for the axial current 
\be
\left\{
\begin{array}{c}
\calM_t[A] \\
\calM_\lambda[A]
\end{array}
\right\}
= 
\left\{
\begin{array}{c}
\calM_t[\hat A] \\
\calM_\lambda[\hat A] 
\end{array}
\right\}
+
\frac{1}{m_\pi^2 - t}
\left\{
\begin{array}{c}
\omega \\
\dlz |\vq|
\end{array}
\right\}
 \left(
  \omega \calM_t[\hat A]
  - |\vq| \calM_0[\hat A]
  + \calM[\hat P]\right),
\ee
\be
\calM_t[\hat A] &=&
\sqrt2 g_A\, \int_0^\infty\! dr\, \left\{
\frac{1}{m_N} 
 u_0^{(1)} (u_d - \sqrt2 w_d) j_1
- \frac{q \omega}{8 m_N^2}
  u_0 (u_d j_0 + \sqrt2 w_d j_2)
\right\}
\nonumber \\
&-& \frac{\sqrt2 \; g_A}{f_\pi^2}
\left[1 - \left(\hat c_2 + \hat c_3 - \frac{g_A^2}{8}\right)
 \frac{m_\mu - E_\nu}{m_N} \right]
\int_0^\infty\! dr\, 
 u_0 (u_d - \sqrt2 \; w_d) j_1\,\frac{y_{1\Lambda}}{r},
\\
\calM_\lambda[\hat A] &=&
\sqrt2 \; g_A\, \int_0^\infty\! dr\, \left\{
\left[ 1 + \frac{t}{6} r_A^2 - \frac{\vbp^2}{3 m_N^2}
 - \dlz \frac{\vq^2}{8 m_N^2} \right]
 u_0 \left( u_d  j_0 - \frac{w_d}{\sqrt2} j_2^\lambda \right)
\right.
\nonumber \\
&& \left.
- \frac{1}{6 m_N^2} u_0^{(2)}
 \left[ \left(u_d - \frac{w_d}
 {\sqrt2}\right) j_2^\lambda - \sqrt2 
  w_d j_0\right]
\right\}
\nonumber \\
&-&  (4\sqrt2) \frac{g_A}{2 m_N f_\pi^2}\,
 \int_0^\infty dr\, \left[
\frac{y_1}{r} \left( \calO^{\rm kin}
- \frac{1+c_6}{4} (1-\dlz) |\vq| u_0 
- (u_d + \frac{w_d}{\sqrt2}) j_1
\right)
\right.
\nonumber \\
&&-\frac{\mpit^2}{3} y_{0\Lambda} 
 \left(\hat c_3 + 2 \hat c_4 + \frac12\right)
 u_0 \left( u_d \; j_0 - 
 \frac{w_d}{\sqrt2} j_2^\lambda\right) 
\nonumber \\
&&+ \frac{\mpit^2}{3} y_{2\Lambda} 
 \left(\hat c_3 - \hat c_4 - \frac14\right)
 u_0 \left( \sqrt2 \; w_d \; j_0 - 
(u_d - \frac{w_d}{\sqrt2}) j_2^\lambda\right) 
\nonumber \\
&&\left.+ \hat d^R\, \delta_\Lambda(\vr) u_0 u_d
 \frac{}{} \right] \, , 
\label{axialtwobodyme} 
\\
\calM[\hat P ] &=& (4\sqrt2) \frac{g_A}{2 m_N f_\pi^2}\,
 \int_0^\infty dr\, \left[ 
  2 \hat c_1 m_\pi^2
 \frac{y_{1\Lambda}}{r} u_0 (u_d - \sqrt2 w_d) j_1 \right] \, ,
\ee
where
\be
\calO^{\rm kin} &=&
 -\dlz \frac{|\vq|}{8} u_0 (u_d - \sqrt2 \; w_d) j_1
\nonumber \\
&+& \frac{1}{12} (j_0 + j_2^\lambda)
 \left[ u_0 (u_d' - \sqrt2 \; w_d')
 - u_0' (u_d - \sqrt2 \; w_d) \right]
-\frac{1}{4\sqrt{2}} (2 j_0 - j_2^\lambda) u_0 w_d \;,
\ee
$j_2^\lambda\equiv (1-3 \dlz) 
j_2$, $j_n = j_n  (\frac12 q r )$,
$u_0^{(1)}(r) = u_0'(r) - 
\frac{u_0(r)}{r}$ and
$u^{(2)}(r) = u_0''(r) - 
3 \frac{u_0'(r)}{r} + 3 \frac{u_0(r)}{r^2}$.
In the above expressions 
the curly brackets denote 1B contributions,
and for clarity we have suppressed the 
dependence on $r$ in some equations. 

\section{Results}
\indent

Table~\ref{total} shows $\Gamma_{\mu d}^{L=0}$
as a function of the cut-off parameter, $\Lambda$. 
As discussed, the short-range exchange current contribution 
depends on the single low-energy constant $\dR$, 
see Eqs. (\ref{axialtwobodyme}), (\ref{dR}), 
and $\dR$ determined from $\Gamma_\beta^t({\rm exp})$
is a function of $\Lambda$ (see Ref.\cite{pp-hep}). 
\begin{table}[hbt]
\begin{center}
\begin{tabular}{|c|c|c|} \hline
$\Lambda$ (MeV) & $\dR$ & $\Gamma_{\mu d}^{L=0}$ $[s^{-1}]$  \\ \hline
500 & $1.00\pm 0.07$ & $254.7 - 9.85\ \dR + 0.159\ (\dR)^2=245.0\pm 0.7$ \\
600 & $1.78\pm 0.08$ & $261.1 - 9.09\ \dR + 0.132\ (\dR)^2=245.3\pm 0.7$ \\
800 & $3.90\pm 0.10$ & $271.0 - 6.76\ \dR + 0.070\ (\dR)^2=245.7\pm 0.6$ \\
\hline
\end{tabular}
\caption{\label{total}\protect
$L=0$ capture rate (in $s^{-1}$)
calculated as a function of the cutoff $\Lambda$.
Also listed are the corresponding values 
of $\dR$ determined from $\Gamma_\beta^t({\rm exp})$
\cite{pp-hep}.}
\end{center}
\end{table}
We observe that
the variation of $\Gamma_{\mu d}^{L=0}$
over the range of $\Lambda$ under consideration
is less than $0.7\ s^{-1}$.
The $\dR$-dependence in the table
indicates the importance of the 
contribution of the short-distance exchange current. 
Without the $\dR$ term, 
$\Gamma_{\mu d}^{L=0}$ would change as much as
$16\ s^{-1}$ for $\Lambda$ = 500 - 800 MeV.
Thus, renormalizing the $\dR$-term
using $\Gamma_\beta^t({\rm exp})$
reduces the variation of $\Gamma_{\mu d}$
with respect to $\Lambda$
by a factor $\approx 20$, 
leading to the practically $\Lambda$-independent 
behavior of $\Gamma_{\mu d}$.
Considering this stability we will hereafter 
only discuss the case corresponding to $\Lambda=600$ MeV
and $\dR=1.78$.

The capture rate contains several interference terms,
which are listed in Table~\ref{cumul}
in a cumulative manner.
\begin{table}[hbt]
\begin{center}
\begin{tabular}{|c|ccc|c||c|} \hline
$\Gamma_{\mu d}$ $[s^{-1}]$ &  & $L=0$ &  & $L\ge 1$ & Total\\
   &  $|\mbox{GT}|^2$& $|\mbox{GT+AC}|^2$& $|\mbox{GT+AC+M1}|^2$&  & \\ \hline
$\left|\mbox{   1B}\right|^2$ & 178 & 177 & 232 & 138 & 370 \\ \hline
$\left|\mbox{1B+2B}\right|^2$ & 187 & 186 & 245 & 141 & 386 \\ \hline
\end{tabular}
\caption{\label{cumul}\protect
Cumulative contributions to $\Gamma_{\mu d}$
(calculated for $\Lambda=600$ MeV and $\dR=1.78$).
The row labeled ``1B" corresponds to the case
that contains one-body contributions only,
while the row labeled ``1B+2B" to the case
that includes both the one-body and MEC contributions.
The three columns labeled  ``L=0" show contributions from 
the $L=0$ channel, with the contributions
of the different transition operators 
displayed in a cumulative manner.
The fifth column gives contribution
from the $L\ge 1$ channels, as evaluated in TKK, Ref.\cite{tkk},
and the last column shows the sum of 
the $L=0$ and $L\ge 1$ contributions.}
\end{center}
\end{table}
We note that the axial charge (AC)  
plays only a minor role;
its destructive interference with GT decreases the capture
rate by $\sim 1\ s^{-1}$.
Meanwhile, the M1 contribution interferes constructively 
with GT, increasing $\Gamma_{\mu d}$
by $\sim 59\ s^{-1}$.
Furthermore, the two-body MEC in the $L=0$ channel
increases the capture rate 
by $\sim 13\ s^{-1}$.

Our final result for $\Gamma_{\mu d}^{L=0}=245\ s^{-1}$ 
in Table~\ref{cumul}
should be compared with 
TKK's result, $\Gamma_{\mu d}^{L=0}$(TKK) 
= 259 $s^{-1}$.\footnote{
We have re-run the code of TKK using $g_A=1.267$.
TKK's original result corresponding to $g_A=1.262$
was $\Gamma_{\mu d}^{L=0}$(TKK)$=257\ s^{-1}$.}
By adding the $1\le L \le 5$ contribution,
$\Gamma_{\mu d}^{L\ge 1}=141\ s^{-1}$, calculated by TKK,
we arrive at the total capture rate
\be
\Gamma_{\mu d}= 386\ s^{-1} , 
\ee
to be compared 
with TKK's result 
$\Gamma_{\mu d}({\rm TKK})=(397 \sim 400)\ s^{-1}$. 

\begin{figure}[htb]
\begin{center}
\epsfig{file=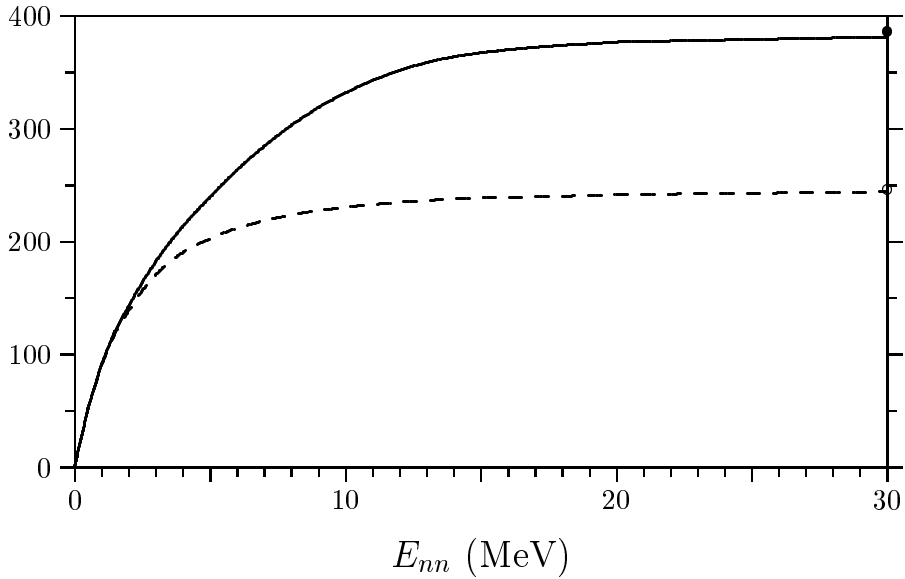}
\caption[]{\label{caprate} \protect
Cumulative $\mu$-$d$ capture rate (in $s^{-1}$) 
calculated for $\Lambda=600\ \MeV$ and $\dR=1.78$.
The dashed line gives the $L=0$ contribution, 
$\Gamma_{\mu d}^{L=0}(E_{nn})$,
while the solid line shows the total contribution,
$\Gamma_{\mu d}(E_{nn}) \equiv
\Gamma_{\mu d}^{L=0}(E_{nn})
+\Gamma_{\mu d}^{L\ge 1}(E_{nn})$.
The empty and solid circles for the
values at $E_{nn}= E_{nn}^{\Max}\simeq 102\ \MeV$,
for $L=0$ and $L\ge 0$, respectively.}
\end{center}
\end{figure}

As mentioned earlier,
a primary question is whether $\mu d$ capture process 
is ``gentle enough" for applying HB$\chi$PT  
with reasonable confidence. 
As noted the ``dangerous" region for HB$\chi$PT occurs  
when the two neutrons carry most of the final energy.
To address this issue,
it is useful to consider the differential capture rate,
$d\Gamma_{\mu d}/dE_{nn}$, 
where $E_{nn}\equiv 2 (\sqrt{m_n^2 +\vp^2} - m_n)$
is the energy of the final two-neutron relative motion.
An equally informative quantity
is the ``cumulative" capture rate
\be
\Gamma_{\mu d}(E_{nn})\equiv
\int_0^{E_{nn}}\!(d\Gamma_{\mu d}/dE_{nn}')\ dE_{nn}'.
\ee
From these quantities we can assess
to what extent $\mu d$ capture is free from 
the ``dangerous" kinematic region. 
\begin{table}[htb]
\begin{center}
\begin{tabular}{|c|cccc|r|}
\hline
$E_{nn}$
  & $\calM_{+1}[A]$
  & $\calM_{+1}[V]$
  & $\calM_{ 0}[A]$
  & $\calM_{ t}[A]$
  & $\Gamma_{\mu d}^{L=0}$ $[s^{-1}]$ \\ \hline
  0.0 & $73.09 + 1.24$ &$14.68 + 0.53$ & $50.22 + 0.81$ & $0.79 -0.23$
& 0 \\
  1.0 & $20.88 + 0.38$ & $4.15 + 0.16$ & $14.26 + 0.25$ & $0.18 -0.07$
& 91 \\
 10.0 &  $2.59 + 0.12$ & $0.47 + 0.04$ &  $1.82 + 0.08$ & $0.06 -0.01$
& 231 \\
 30.0 &  $0.49 + 0.05$ & $0.07 + 0.01$ &  $0.39 + 0.04$ & $0.04 -0.00$
& 244 \\
$E_{nn}^{\Max}$
&  $0.056 -0.003$& 0 &  $0.056 -0.003$ & 0
& 245 \\
\hline
\end{tabular}
\caption{\label{tabmes}\protect
Matrix elements calculated for representative values
of $E_{nn}$ (MeV) and the cumulative $L=0$ capture rate
for the case: $\Lambda=600\ \MeV$ and $\dR=1.78$.
In each entry for the matrix element, 
the first number
(preceding a ``+" or ``$-$" sign)
gives the one-body contribution, 
while the second number gives 
the two-body contribution.}
\end{center}
\end{table}
We show in {}Table \ref{tabmes} the matrix elements,
$\calM_{+1}[A]$, $\calM_{+1}[V]$, $\calM_{ 0}[A]$  
and $\calM_{ t}[A]$, calculated for representative values
of $E_{nn}$, 
and for $\Lambda=600\ \mbox{MeV}$ and $\hat d^R=1.78$.  
{}Table \ref{tabmes} also gives $\Gamma^{L=0}_{\mu d}(E_{nn})$. 
The graphical representation of $\Gamma_{\mu d}(E_{nn})$
can be found in Fig. \ref{caprate}.
%-- place for figure
We learn from {}Table \ref{tabmes} that
the matrix elements decrease quite fast as $E_{nn}$ increases,
a feature that can be easily understood as follows. 
The ${}^1S_0$ $nn$ radial wave function 
is proportional to  
$\frac{\sin\delta_0}{p} = 
\pm \left[(p\cot\delta_0)^2 + p^2\right]^{-\frac12}$.
Since the $nn$ scattering length is very large, 
$p\cot\delta_0$ diminishes rapidly
when the $nn$ relative momentum $p$ gets small. 
The examination of Table \ref{tabmes} also reveals
that the 
one-body amplitudes decrease more quickly than the
two-body amplitudes. 
This is a consequence of the softness of the 
deuteron wave function, which cannot supply  
high momentum transfers needed for producing 
large values of $p$.
As a result, the contributions from high $E_{nn}$ 
-- where the applicability of EFT is questionable --
is negligible. 
For instance,
the contribution to $\Gamma_{\mu d}^{L=0}$
from $E_{nn} > 30$ MeV is 
just $1.1\ s^{-1}$, and that from
$E_{nn} > 50$ MeV is less than $0.1\ s^{-1}$. 

We now can make a rough estimate of the 
theoretical error associated with this calculation.
Uncertainty related to the $G_P$ term, 
Eq.(\ref{eq:GP}) (or $\beta$)
is $\sim 1\ s^{-1}$, while 
uncertainty reflecting the $\Lambda$-dependence 
is less than $1\ s^{-1}$;
uncertainty in $\Gamma_\beta^t({\rm exp})$
(or that in $\dR$ for a given $\Lambda$) 
can affect $\Gamma_{\mu d}$
at the level of $1\ s^{-1}$.
If we assign a rather conservative error, 
$2\ s^{-1}$, to the $L\ge 1$ contributions 
obtained in Ref. \cite{tkk},
the overall uncertainty in our estimate 
becomes $4\ s^{-1}$
or $1\ \%$ in the total capture rate.

As mentioned, there is a serious disagreement between 
the two measured values of $\Gamma_{\mu d}$.
Our theoretical result is consistent with 
$\Gamma_{\mu d}(exp)$ in Ref.\cite{caretal89}.
In the present exploratory study
we have not considered 
radiative corrections \cite{bp01},
which are expected to be smaller than 
the existing uncertainty in $\Gamma_{\mu d}(exp)$.
When the planned precision measurement of the 
$\Gamma_{\mu d}$ at PSI \cite{kametal} is realized,
the issue of radiative corrections
should certainly be addressed.
The EFT approach as described here
will provide a useful tool for this purpose as well.
Once the accuracy in $\Gamma_{\mu d}(exp)$
is significantly improved,
we will be able to use $\mu d$ capture
to determine the low energy constant $\dR$, 
a quantity critically important 
for the accurate evaluation of the 
$\nu d$ cross sections used in the analysis  
of the SNO experiments. 
At present the tritium $\beta$-decay
is a much more accurate source of information 
on $\dR$ than $\mu d$ capture,
but it is hoped that in the near future
$\Gamma_{\mu d}$ will provide an independent constraint
on $\dR$.
We consider this redundancy extremely important.

We thank P. Kammel and J.-W. Chen for useful communications.
This work is supported in part by NSF Grant No. PHY-9900756
and No. INT-9730847.

\end{document}